\documentstyle[psfig,12pt,aas2pp4]{article}
\begin{document}
\title{Discovery of kilohertz quasi-periodic oscillations in the
Z source GX\,340+0}

\author{Peter G. Jonker\altaffilmark{1}, Rudy
Wijnands\altaffilmark{1}, Michiel van der Klis\altaffilmark{1},
Dimitrios Psaltis\altaffilmark{2}, Erik
Kuulkers\altaffilmark{3}, Frederick K. Lamb\altaffilmark{4}}
\altaffiltext{1}{Astronomical Institute ``Anton Pannekoek'', University of
Amsterdam, and Center for High-Energy Astrophysics, Kruislaan 403,
1098 SJ Amsterdam; peterj@astro.uva.nl, rudy@astro.uva.nl, 
michiel@astro.uva.nl}
\altaffiltext{2}{Harvard-Smithsonian Center for Astrophysics, 60
Garden Street, Cambridge, MA 02138, USA; demetris@cfata1.harvard.edu}
\altaffiltext{3}{Astrophysics, University of Oxford, Nuclear and
Astrophysics Laboratory, Keble Road, Oxford OX1 3RH, United Kingdom;
e.kuulkers1@physics.oxford.ac.uk}
\altaffiltext{4}{Department of Physics and Astronomy, University of
Illinois at Urbana-Champaign, Urbana, IL 61801; f-lamb@uiuc.edu}

\begin{abstract}
We have discovered two simultaneous kHz quasi-periodic oscillations 
(QPOs) in the Z source GX\,340+0 with the Rossi X-ray Timing
Explorer. The X-ray
hardness-intensity and color-color diagram each show a full Z-track,
with an extra limb branching off the flaring branch of the Z. 
Both peaks moved to higher
frequencies when the mass accretion rate increased. The two peaks moved
from $247 \pm 6$ and $567 \pm 39$ Hz at the left end of the horizontal
branch to $625 \pm 18$ and $820 \pm 19$ Hz at its right end. The
higher frequency peak's rms amplitude (5--60 keV) and FWHM decreased from
$\sim 5\%$ and $383 \pm 135$ Hz
to $\sim 2\%$, and $145 \pm 62$ Hz, respectively. The rms
amplitude and FWHM of the lower peak were consistent with being constant
near 2.5 \% and 100 Hz. The kHz QPO separation was 
consistent with being constant at $325 \pm 10$ Hz.
\newline
\indent
Simultaneous with the kHz QPOs we detected the horizontal branch
oscillations (HBO) and its second harmonic, at frequencies between 20
and 50 Hz, and 38 and 69 Hz, respectively.
The normal branch oscillations were only detected on the upper and
middle normal branch, and became undetectable on the lower
normal branch. The HBO frequencies do not
fall within the range predicted for Lense-Thirring (LT) precession, 
unless either the ratio of the
neutron star moment of inertia to neutron star mass is at least 4, 
$10^{45}g\,cm^2/M_{\odot}$, the frequencies of the HBO 
are in fact the sub-harmonic oscillations, or the observed kHz peak
difference is half the spin frequency and not the spin frequency.
\newline
\indent
During a 1.2 day gap between two observations, the Z-track in the
hardness-intensity diagram moved to higher
count rates by about 3.5\%. Comparing data before and after this
shift, we find that the HBO properties are
determined by position on the Z-track and not
directly by count rate or X-ray colors.
\end{abstract}

\section{Introduction}
GX\,340+0 is a bright low-mass X-ray binary and a Z source
(\cite{hasinger}). 
The Z-track
traced out in the X-ray color-color (CD) and hardness-intensity
diagram (HID) is
usually built up of three
components, the horizontal branch (HB), the normal branch (NB), and the
flaring branch (FB). The mass accretion rate ($\dot{M}$) is thought to increase
from a few tens of the Eddington mass accretion rate when the source is 
on the HB, to near-Eddington on the
NB, and super-Eddington on the FB. On the HB and the
upper part of the NB of GX\,340+0, quasi-periodic oscillations (QPOs) with
frequencies of 32--50 Hz, the horizontal branch oscillations
(HBOs), were detected by Penninx et al. (1991). On the central part of the NB, 
QPOs with a typical 
frequency of 5.6 Hz, the normal branch oscillations (NBOs), 
were detected by van Paradijs et al. (1988). 
\newline
\indent
In four Z sources, Sco\,X-1 (\cite{vanderKlis1996a}; \cite{michiel1997b}),
GX\,5-1 (\cite{michiel1996b}), GX\,17+2 (\cite{wijnands1997b}), and Cyg\,X-2
(\cite{wijnands1997c}) two simultaneous kilo Hertz (kHz) QPOs have
been observed, with frequencies reflecting flux
changes on timescales close to the dynamical timescales near a
neutron star. The higher frequency peak (henceforth called upper peak)
has been proposed to
reflect the Keplerian frequency of blobs of material orbiting around
the neutron star, at a preferred radius in the disk, such as the sonic radius
(Miller, Lamb \& Psaltis 1998). The
lower frequency peak (henceforth called lower peak) is the proposed beat 
frequency between the frequency of the upper peak and the spin frequency of
the neutron star. The kHz QPO
frequency is observed to increase with increasing
$\dot{M}$. Similar kHz QPOs
are also found in atoll sources. 
The frequencies of the kHz QPOs are between 300 and 1200
Hz. (See van der Klis (1997) for a recent overview).
\newline
\indent
In this Letter we report the discovery of two simultaneous kHz QPOs
in the Z source GX\,340+0. A preliminary announcement of these results
was already made by Wijnands \& van der Klis (1998b).

\section{Observations and Analysis}
We observed GX\,340+0 with the proportional counter array onboard
the Rossi X-ray Timing Explorer on 1997 Apr.\,17, Sept.\,21,\,23,\,25,
and Nov.\,1,\,2,\,3, and 4. The total amount of
good data obtained was 138 ksec. During 5\% of this time only three or four of
the five detectors were active.  
For our CD and HID we used only the data from the three
detectors that were always active. In our power spectral analysis we
used all available data. The mean five-detector countrate was $6443$
$\rm{cts\,s^{-1}}$.
The data were obtained using 16s time resolution in
129 photon energy bands (effective energy range 2--60 keV), and simultaneously
with 122 $\mu s$ time resolution in four bands (2--5.0 keV,
5.0--6.4 keV, 6.4--8.6 keV, and 8.6--60 keV) on Apr.\,17 or in two
bands for the other dates (2--5.0 keV and 5.0--60 keV).
\newline
\indent
We calculated power density spectra using the 122 $\mu s$ data divided into
16s segments. To determine the properties of the kHz QPOs we fitted
the 95--4096 Hz power spectra with a function that consisted
of one or two Lorentzians (the kHz QPOs) and a constant plus a broad
sinusoid (the dead-time
modified Poisson noise) (\cite{Zhang2}). 
The Very Large Event window (\cite{michiel1997b}) was set to 55 $\mu s$, 
so that its effect on the Poisson noise is 
small and could be incorporated in the sinusoid.
To determine the properties of the HBO, we fitted the 0.125--126.5 Hz
power spectra using a function which consisted of the sum of 
a constant, one or two Lorentzian peaks, and two exponentially cut-off
power-law components. The errors were determined using  $\Delta \chi^2
=1.0$ (1\,$\sigma$ single parameter) and upper
limits using $\Delta \chi^2=2.71$, corresponding to a $95\% $ confidence level.
Upper limits on the kHz QPOs were determined using a fixed FWHM of 150
Hz. When only the lower peak was detected, the upper limit on the
upper peak was determined by setting the frequency to the value of
the frequency of the lower peak
plus the mean frequency difference between the two peaks. Upper limits on the
HBO were determined by keeping the FWHM fixed at 10 Hz.
\newline
\indent
In the HID and CD, the hard color
is defined as the logarithm of the 9.7--16.0 keV/6.4--9.7 keV countrate
ratio. The soft color is defined as the logarithm of the 3.5--6.4 keV/2--3.5
keV countrate ratio.
The intensity is defined as the logarithm of the countrate in the 2--16.0 keV
band. The HID and CD are background-corrected but no dead-time
correction was carried out. The dead-time correction factor was $\sim 2\%$.
In order to measure the position along the Z-track, we use the $S_z$ 
parametrization
(e.g. \cite{wijn1997a}) applied to the HID. In this parametrization the HB/NB 
vertex is fixed at $S_z =1.0$ and the NB/FB vertex at $S_z =2.0$. We 
selected the power spectra according to $S_z$, and determined the 
average $S_z$ and standard deviation for each selection.
\newline
\indent
The track of the Nov.\,4 data is shifted towards higher intensity
and softer colors with respect to the Z-track in the other data. We 
therefore determined the HBO properties for 
the Nov.\,4 data ($\sim$ 18 ksec) separately. To
determine the kHz QPO properties we combined the Nov.\,4 data with
the rest in order to get a higher signal to noise ratio.
As the kHz QPOs were only significantly detected in the 5.0--60 keV
band and the significance did not increase combining the two
bands, we used only the 5.0--60 keV band in our analysis. 

\section{Results}
We observed a full Z-track in the HID and CD (Fig. \ref{HID+CD}).
The Z-shape shows a fourth branch beyond the FB. This extra,
trailing branch was
seen before in GX\,340+0 (\cite{penninx}) and GX\,5-1
(\cite{kuulkersetal}; \cite{rudyconfproc}).
Detailed examination of the HB/NB vertex in the HID revealed that the
1997 Nov.\,4 data were shifted to higher count rates ($\sim
3.5\%$) and softer colors ($\sim 0.5\%$) than the other 
data. Due to this shift, which occurred during a 1.2 day gap
between two observations, the HB/NB vertex in the HID is
broadened. Due to the size of the dots in 
Fig. \ref{HID+CD} this is not visible. This shift was also
seen in EXOSAT data of GX\,340+0 (\cite{kuulkers}).
\newline
\indent
We detected kHz QPOs in the 5.0--60 keV band (Fig. \ref{PDSkHz+HBO}).  
The lower kHz peak was detected at $S_z$ values up to 1.04, the 
upper peak up to $S_z=0.95$ (Fig. \ref{kHz+HBO}). 
The lower peak had a frequency of $247 \pm 10$ Hz at
$S_z=0.47$ and increased to $625 \pm 18$ Hz at
$S_z=1.04$, its rms amplitude remained about constant
at values near 2\%, and its FWHM varied between 30 and 160 Hz with no 
clear correlation with $S_z$.
The frequency of the upper peak was $567 \pm 39$ Hz at  $S_z=0.47$ and 
increased to $820 \pm 19$ Hz at $S_z=0.95$, while its rms
amplitude decreased from $5.1 \pm 0.9 \%$ to $2.0 \pm
0.4 \%$, and its FWHM from $383 \pm 135$ Hz to $145 \pm 62$ Hz.
The FWHM ratio did not significantly depend on $S_z$. 
When both peaks were detected, the peak separation was consistent with
being constant at a value of $325 \pm 10$ Hz. We did not detect 
significant kHz QPOs in the 2--5.0 keV bands, with upper limits of 2.8\% and
2.3\% at $S_z=0.58$ and 2.7\% and 2.8\% at $S_z=0.87$ for the lower
and upper peaks, respectively, while in the 5.0--60 keV band the peaks
were found with a rms amplitude of 1.5\% and 4.4\% at $S_z=0.58$
and 1.6\% and 2.8\% at $S_z=0.87$, respectively.
\newline
\indent
Simultaneously with the kHz QPO, the HBO and its second harmonic were
detected. When we used
the same $S_z$ selections as in the fitting of the kHz QPOs the HBO
sometimes showed double peaks, especially at low $S_z$. This was
caused by the steep dependence of the HBO
frequency on $S_z$ and its narrow peak width. Therefore up
to $S_z=0.7$, we used narrower $S_z$
selections in fitting the HBO to prevent the HBO peak from moving too
much within the selection.
The HBO frequency increased from $19.43 \pm 0.03$ Hz at $S_z=0.43$ to 
$49.92 \pm 0.21$ Hz at $S_z=1.05$. Then it remained constant up to
$S_z=1.25$ (Fig. \ref{kHz+HBO} b). 
The rms amplitude in the 5.0--60 keV energy band of the HBO decreased
smoothly from $9.1 \pm 0.1$\% at the left side of the HB ($S_z=0.43$) to
$1.8 \pm 0.2$\% on the NB ($S_z=1.25$). For higher $S_z$
selections, no HBO was detected, with upper limits of $\sim 2\%$
rms. 
The FWHM increased from $2.9 \pm 0.1$ Hz to $15.3 \pm 4.2$ Hz. 
We did not find a clear dependence on $S_z$ 
of the rms amplitude and FWHM of the second harmonic. Due to the
presence of a broad noise component around the frequency of the second
harmonic it was difficult to determine the properties of either one of the 
two components.
\newline
\indent
We found that the HBO frequency vs. $S_z$ relation of the Nov.\,4
data was offset with respect to that of the other data
(Fig. \ref{HBO+4novHBO} left), due to 
the shift in the vertex of the Nov.\,4 data, which
along the HB, corresponds to a shift in $S_z$  
of $\sim 0.05$, equal to the offset in the $S_z$
vs. $\nu_{HBO}$ plot (Fig. \ref{HBO+4novHBO}, left).
If we correct for the change in $S_z$ by measuring $S_z$ on Nov.\,4 
from the HB/NB vertex appropriate for that date,
the HBO frequencies, rms amplitude, and FWHM of the Nov.\,4 data 
and the HBO frequencies, rms amplitude, and FWHM of the other data as 
a function of $S_z$ are the same
(Fig. \ref{HBO+4novHBO}, right). This strengthens previous conclusions
(\cite{kuulkersetal}) that $S_z$, not count rate, is the better
measure for $\dot{M}$.
\newline
\indent  
We also detected the NBOs, between $S_z \sim 1.1$ and $S_z \sim
1.8$. These will be more extensively discussed elsewhere.

\section{Discussion}
We have discovered kHz QPOs in the Z source GX\,340+0. The frequencies of
both the lower and upper peak increased as the source
moved along the HB to the HB/NB vertex. The peak separation was
consistent with being constant.
This is similar to
what was found in GX\,17+2 (\cite{wijnands1997b}), Cyg\,X-2 
(\cite{wijnands1997c}), and GX\,5-1 (\cite{michiel1996b}; 
\cite{wijnands1998c}). Only in Sco\,X-1 
(\cite{michiel1997b}) and perhaps in the atoll source 4U 1608-52
(\cite{mendez1998}) the peak separation was found to decrease with
increasing $\dot{M}$. If this 
decrease in Sco X-1 is in some way related to the source approching the 
Eddington critical luminosity (\cite{white1997}; \cite{milleretal}), 
this could mean that if we were
able to detect the kHz QPOs in GX\,340+0 and the other Z-sources
(further) up the NB we might also see the peak separation decrease.
\newline 
\indent
The lower peak reaches the so far lowest ``kHz'' QPO frequency 
found in any low mass X-ray binary, 247 Hz.  
The maximum frequency reached by the upper peak, 820 Hz, is also
rather low. Using the sonic-point model (\cite{milleretal}) to
explain the upper peak kHz QPO frequency, we
found lower limits on the mass of the neutron star of $M_{neutr} >
0.6M_{\odot}$ for a neutron star radius of 4M, and  $M_{neutr} >
1.2M_{\odot}$ for a neutron star radius of 7M. (The implicit
assumptions made in calculating these lower limits are: 
Schwarzschild spacetime metric, non-rotating radiating layer, and photons 
remove angular momentum from the disk material only in one scattering event.)
We note that the frequencies of the kHz QPOs seem to confirm the idea
(e.g. \cite{kuulkersetal}) that GX\,340+0 is more similar in
appearance to GX\,5-1 (\cite{wijnands1998c}), than to Sco\,X-1
(\cite{michiel1997b}), and GX\,17+2 (\cite{wijnands1997b})
(frequencies higher than 1000 Hz). 
\newline
\indent
Any straightforward beat-frequency model predicts the upper peak to be
narrower than, or as narrow as the lower peak. We find in GX\,340+0
that the upper peak is usually broader than the lower peak. In fact, 
the upper peaks found in GX\,340+0 were very broad with 
$\frac{\delta \nu}{\nu}$ up to 0.67 at $S_z=0.43$.  
Therefore a straightforward beat-frequency interpretation of the two peaks in 
GX\,340+0 is difficult. In GX\,340+0, GX\,5-1 (\cite{wijnands1998c}), 
and GX\,17+2 (\cite{wijnands1997b}) the upper peak FWHM
decreases as a function of $S_z$. To explain
this behaviour one could involve scattering in a rapidly variable
medium (\cite{mendez1998}). The effect of this scattering should become
less as $\dot{M}$ increases. 
\newline
\indent 
Although the behaviour of the kHz QPOs as a function of inferred 
$\dot{M}$ in GX\,340+0
is similar to that found in other Z sources, the HBO behaviour is
rather different.
The FWHM of the HBO increased as the rms amplitude decreased, which is
unlike what is seen in other Z sources. 
In Cyg\,X-2 Wijnands et al. (1997c) found the FWHM to be
well-correlated to the rms. In GX\,17+2 the 
FWHM remained about 
constant while the rms amplitude decreased (\cite{homan}).
In previous observations of GX\,340+0 (\cite{penninx}, Table 2)
the behaviour is consistent with our findings. In
observations of Kuulkers \& van der Klis (1996) the FWHM does not show a
clear relation as a funtion of the rms amplitude. Exactly the same
behaviour (as we reported for GX\,340+0) was found in GX\,5-1
(\cite{lewin1992}; \cite{wijnands1998c}). This suggests that an extra parameter
in addition to $\dot{M}$ determines the timing properties.
Another indication of this is that there are
differences between the Z sources in the position on the Z-track 
where the NBO are found. In GX\,17+2 (\cite{penninx1990}), and Sco\,X-1
(\cite{vanderKlis1987}) the NBO were detected up the FB, e.g. $S_z > 2$.
In Cyg\,X-2 they were detected up to the NB/FB vertex (\cite{wijn1997a}),
while in GX\,340+0 and GX\,5-1 the NBO were
detected up to $S_z=1.8$.
\newline
\indent
Recently it has been proposed that the HBO frequency may reflect the
general relativistic Lense-Thirring (LT) precession frequency at the inner edge
of a warped accretion disk (\cite{stella}). If the upper peak is 
interpreted as a Keplerian frequency in the inner part of the disk, there is a
quadratic relation between the LT precession frequency and
the upper peak frequency. 
Applying this theory to our observations and assuming the peak
difference to be the neutron star spin frequency, the LT
predictions for the HBO frequency are too low by at least a factor of
2. The same discrepancy by a factor of 2 has been reported for the Z
source GX\,17+2 (\cite{stella}). There are several ways out, but each
appears to create serious difficulties. $I/M$ (with I the
moment of inertia of the neutron star and M its mass) may be a factor of 
2 higher than presently thought, however, no sensible equation of state
appears to be able to produce such values of $I/M$. The kHz QPO peak 
difference may not
be equal to the neutron star spin frequency but to half the spin
frequency, however, at least in atoll sources this seems to contradict
the results obtained from burst QPO (\cite{strohmayer}). The HBO
may not be the fundamental frequency but the harmonic of the predicted
LT frequency, however, in none of the Z sources there is any evidence
of excess power at frequencies half the HBO frequencies.

\acknowledgments
We would like to thank Jan van Paradijs for commenting on 
the manuscipt and Cole Miller for the useful discussion on the
sonic-point model and the implications of our findings.
This work was supported in part by the Netherlands Foundation for
Research in Astronomy (ASTRON) grant 781-76-017. D.P. acknowledges
support from a Postdoctoral fellowship at the Smithsonian Institute 
and F.K.L. (NAG 5-2925) from US NASA grants.

\clearpage

\figcaption{Hardness-intensity (left) and color-color diagram (right)
of GX\,340+0. The hard color is defined as the logarithm of the
9.7--16.0 keV / 6.4--9.7 keV count rate ratio and the soft color as the 
logarithm of the 3.5--6.4 keV/ 2--3.5 keV countrate ratio. The three-detector 
countrate is measured in the 2--16.0 keV band. The data was background
subtracted but no dead-time correction was applied. The typical
error bars are visible in the HID and CD in points in the extra
trailing branch.
\label{HID+CD}}

\figcaption{Typical power density spectra of GX\,340+0; the power is in units 
of fractional amplitude squared per Hertz. The kHz QPOs are at 
frequencies of 452 and 753 Hz, respectively, at $S_z=0.87$. The rise
in the power density of the left plot towards higher frequencies is caused by 
instrumental dead-time effects on the Poisson noise.
\label{PDSkHz+HBO}}

\figcaption{(A) Frequencies of the kHz QPOs, (B) frequencies of the HBO and
its harmonic, (C) rms amplitude of the upper kHz QPO, (D) rms amplitude
of the HBO, (E) rms amplitude of the
lower kHz QPO, and (F) FWHM of the HBO, as a function
of $S_z$. The error bars on $S_z$ are standard deviations.
\label{kHz+HBO}}

\figcaption{Left: The HBO frequencies of the November\,4 data
(circles) and the other data (X-es). Right: The frequencies of the HBO
for the November\,4 data corrected for
the shift (circles) and the other data (X-es). The
error bars in the frequency are smaller than the size of the symbols.
(see also Section 3 paragraph 4).
\label{HBO+4novHBO}}

\clearpage
\begin{figure*}
\centerline{\psfig{figure=HID+cd+ERR.ps,angle=-90,width=12cm}}
\end{figure*}

\clearpage
\begin{figure*}
\centerline{\psfig{figure=just_kHzfit.ps,width=10cm}}
\end{figure*}

\clearpage
\begin{figure*}[hb]
\centerline{\psfig{figure=kHz+HBOplot.ps,width=12cm}}
\end{figure*}

\clearpage
\begin{figure*}[hb]
\centerline{\psfig{figure=shift+backHBOnu.ps,angle=-90,width=12cm}}
\end{figure*}


\begin{thebibliography}{99}
\bibitem[Hasinger \& van der Klis 1989]{hasinger} Hasinger, G., \&
van der Klis, M. 1989, A\&A, 225, 79
\bibitem[Homan et al. 1998]{homan} Homan, J., et al. 1998, in preparation
\bibitem[Kuulkers et al. 1994]{kuulkersetal} Kuulkers, E., van der
Klis, M., Oosterbroek, T., Asai, K., Dotani, T., van Paradijs, J., \&
Lewin, W.H.G. 1994, A\&A, 289, 795
\bibitem[Kuulkers \& van der Klis 1996]{kuulkers} Kuulkers, E., \& van der
Klis, M. 1996, A\&A, 314, 567
\bibitem[Lewin et al. 1992]{lewin1992} Lewin, W.H.G., Lubin, L.M.,
Tan, J., van der Klis, M., van Paradijs, J., Dotani, T., \& Mitsuda,
K. 1992, ApJ, 256, 545
\bibitem[M\'endez et al. 1998]{mendez1998}  M\'endez, M., et
al. 1998, ApJ, Febr. 20
\bibitem[Miller et al. 1998]{milleretal} Miller, M.C., Lamb, F.K., \&
Psaltis, D. 1998, ApJ, in press
\bibitem[Penninx et al. 1990]{penninx1990} Penninx, W., Lewin, W.H.G.,
Mitsuda, K., van der Klis, M., van Paradijs, J., \& Zijlstra,
A.A. 1990, MNRAS, 243, 114
\bibitem[Penninx et al. 1991]{penninx} Penninx, W., Lewin, W.H.G.,
Tan, J., Mitsuda, K., van der Klis, M., \& van Paradijs, J. 1991, MNRAS,
249, 113
\bibitem[Stella \& Vietri 1998]{stella} Stella, L., \& Vietri, M. 1998,
ApJ, 492, L00
\bibitem[Strohmayer et al. 1996]{strohmayer} Strohmayer, T.E., Zhang,
W., Swank, J.H., Smale, A., Titarchuk, L., \& Day, C. 1996, ApJ, 469, L9
\bibitem[van der Klis et al. 1987]{vanderKlis1987} van der Klis,
M., Stella, L., White, N., Jansen, F., \& Parmar, A.N. 1987, ApJ, 316, 411
\bibitem[van der Klis et al. 1996a]{vanderKlis1996a} van der Klis,
M. Swank, J.H., Zhang, W., Jahoda, K., Morgan, E.H., Lewin, W.H.G., 
Vaughan, B., \& van Paradijs, J. 1996a, ApJ, 469, L1
\bibitem[van der Klis et al. 1996b]{michiel1996b} van der Klis, M., 
Wijnands, R.A.D., van Paradijs, J., Lewin, W.H.G., Lamb, F.K.,
Vaughan, B., Kuulkers, E., Psaltis, D., \& Dieters, S. 1996b, IAU Circ.,
6511   
\bibitem[van der Klis 1997]{michielastroph} van der Klis,
M. Proc. NATO ASI ``The many faces neutron stars'', Lipari, Italy,
1996, astro-ph/9710016, 1997   
\bibitem[van der Klis et al. 1997]{michiel1997b} van der Klis, M.,
Wijnands, R.A.D., Horne, K., \& Chen, W. 1997, ApJ, 481, L97
\bibitem[van Paradijs et al. 1988]{jvp} van Paradijs, J., Hasinger,
G., Lewin, W.H.G., van der Klis, M., Sztajno, M., Schulz, N., \& Jansen,
F. 1988, MNRAS, 231, 379 
\bibitem[White \& Zhang 1997]{white1997} White, N.E., \& Zhang,
W. 1997, ApJ, 490, L87 
\bibitem[Wijnands et al. 1997a]{wijn1997a} Wijnands, R.A.D., van der
Klis, M., Kuulkers, E., Asai, K., \& Hasinger, G. 1997a, A\&A, 323, 399
\bibitem[Wijnands et al. 1997b]{wijnands1997b} Wijnands, R.A.D.,
Homan, J., van der Klis, M., M\'endez, M., Kuulkers, E., van Paradijs,
J., Lewin, W.H.G., Lamb, F.K., Psaltis, D., \& Vaughan, B. 1997b, ApJ,
490, L157  
\bibitem[Wijnands et al. 1998a]{wijnands1997c} Wijnands, R.A.D.,
Homan, J., van der Klis, M., Kuulkers, E., van Paradijs, J., Lewin,
W.H.G., Lamb, F.K., Psaltis, D., \& Vaughan, B. 1998a, in press
\bibitem[Wijnands \& van der Klis 1998b]{rudyconfproc} Wijnands,
R., \& van der Klis, M. 1998b, to appear in "Accretion Processes in
Astrophysical Systems", 
Proc. of the 8th Annual Astrophysics Conference in Maryland, 
S. S. Holt \& T. Kallman (eds.)  
\bibitem[Wijnands et al. 1998c]{wijnands1998c} Wijnands, R., et al. 1998c in preparation
\bibitem[Zhang et al. 1995]{Zhang2} Zhang, W. Jahoda, K., Swank, J.H.,
Morgan, E.H., \& Giles, A.B. 1995, ApJ, 449, 930 
\end{thebibliography}
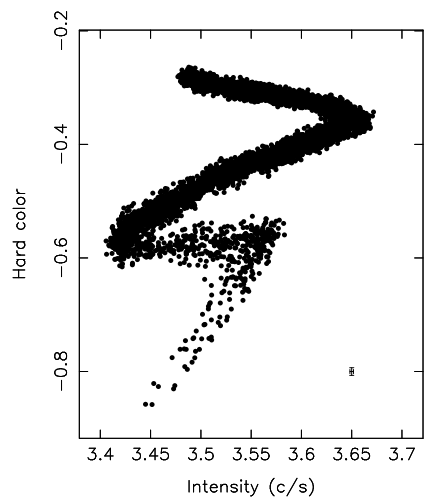
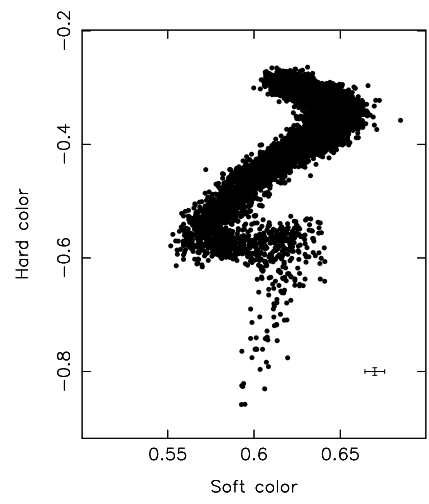
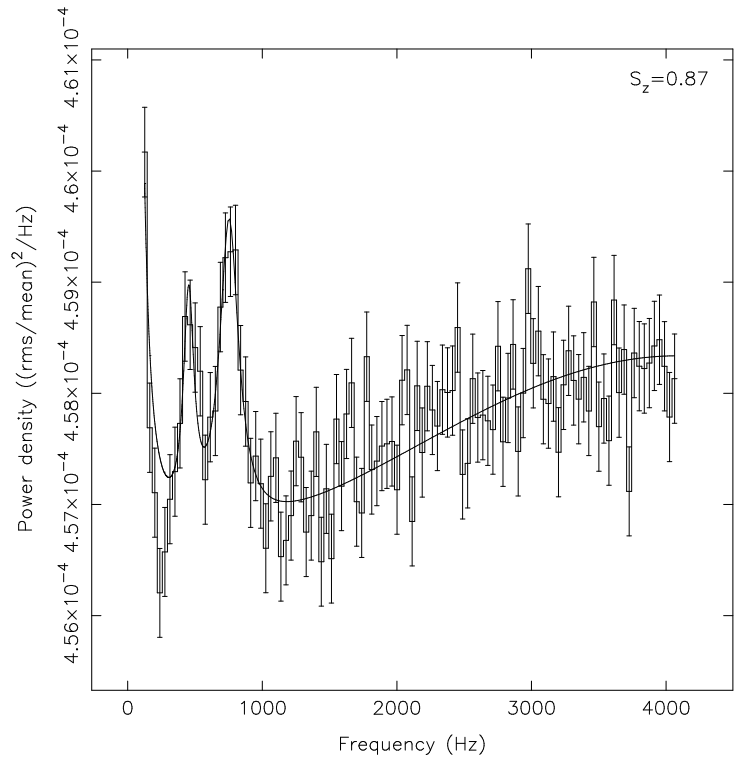
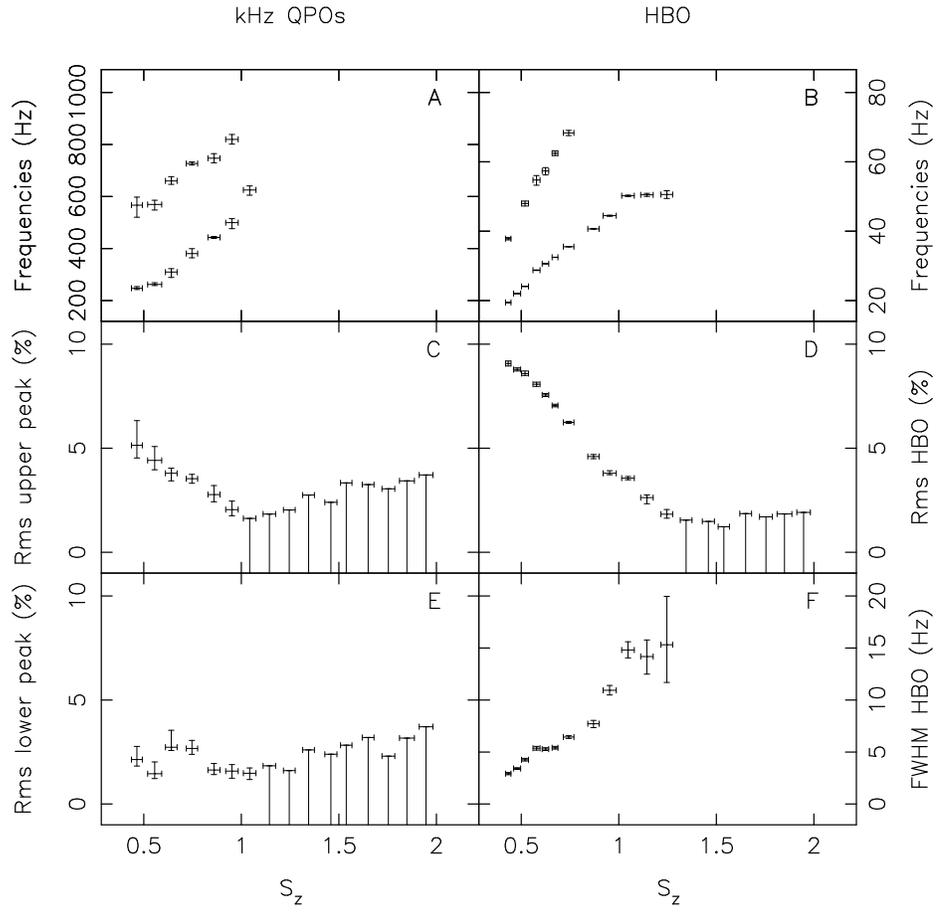
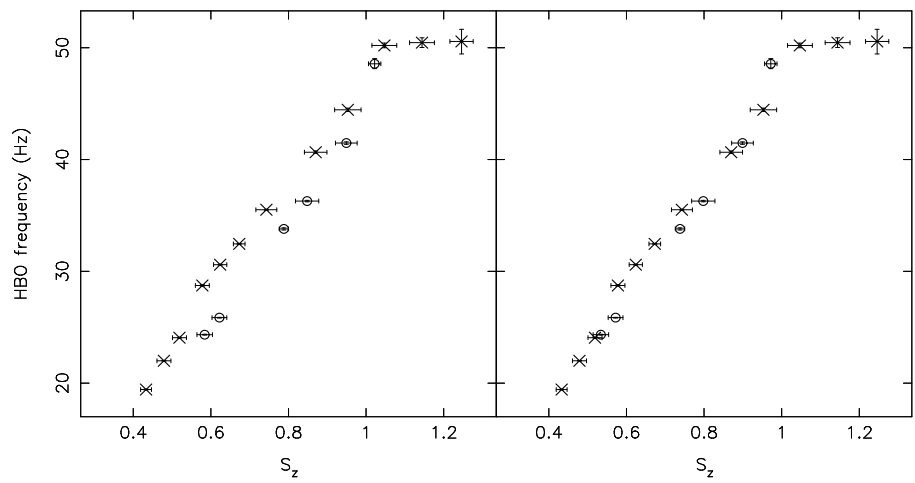
\end{document}